\documentclass{aastex631}

\usepackage{amsmath}

\usepackage{hyperref}

\usepackage{CJK}

\begin{document}

\title{Polarization Position Angle Swing and the Rotating Vector Model of Repeating Fast Radio Bursts}

\begin{CJK*}{UTF8}{gbsn}
\author[0000-0002-2552-7277]{Xiaohui Liu (刘小辉)}
\affiliation{National Astronomical Observatories, Chinese Academy of Sciences, 20A Datun Road, Chaoyang District, Beijing 100101, People's Republic of China}
\affiliation{School of Astronomy and Space Science, University of Chinese Academy of Sciences, Beijing 100049, People's Republic of China}

\author{Heng Xu
(胥恒)}
\affiliation{National Astronomical Observatories, Chinese Academy of Sciences, 20A Datun Road, Chaoyang District, Beijing 100101, People's Republic of China}

\author[0000-0001-8065-4191]{Jiarui Niu (牛佳瑞)}
\affiliation{National Astronomical Observatories, Chinese Academy of Sciences, 20A Datun Road, Chaoyang District, Beijing 100101, People's Republic of China}

\author[0000-0002-8744-3546]{Yongkun Zhang (张永坤)}
\affiliation{National Astronomical Observatories, Chinese Academy of Sciences, 20A Datun Road, Chaoyang District, Beijing 100101, People's Republic of China}
\affiliation{School of Astronomy and Space Science, University of Chinese Academy of Sciences, Beijing 100049, People's Republic of China} 

\author{Jinchen Jiang (姜金辰)}
\affiliation{National Astronomical Observatories, Chinese Academy of Sciences, 20A Datun Road, Chaoyang District, Beijing 100101, People's Republic of China}

\author[0000-0002-6423-6106]{Dejiang Zhou (周德江)}
\affiliation{National Astronomical Observatories, Chinese Academy of Sciences, 20A Datun Road, Chaoyang District, Beijing 100101, People's Republic of China}

\author[0000-0002-9274-3092]{Jinlin Han (韩金林)} 
\affiliation{National Astronomical Observatories, Chinese Academy of Sciences, 20A Datun Road, Chaoyang District, Beijing 100101, People's Republic of China}

\author[0000-0001-5105-4058]{Weiwei Zhu (朱炜玮)} 
\affiliation{National Astronomical Observatories, Chinese Academy of Sciences, 20A Datun Road, Chaoyang District, Beijing 100101, People's Republic of China}

\author{Kejia Lee (李柯伽)} 
\affiliation{National Astronomical Observatories, Chinese Academy of Sciences, 20A Datun Road, Chaoyang District, Beijing 100101, People's Republic of China}
\affiliation{Kavli Institute for Astronomy and Astrophysics, Peking University, Beijing 100871, People's Republic of China}
\affiliation{Department of Astronomy, Peking University, Beijing 100871, People's Republic of China}

\author{Di Li
\begin{CJK}{UTF8}{bsmi}
(李菂)
\end{CJK}}
\affiliation{New Cornerstone Science Laboratory, Department of Astronomy, Tsinghua University, Beijing 100084, People's Republic of China}
\affiliation{National Astronomical Observatories, Chinese Academy of Sciences, 20A Datun Road, Chaoyang District, Beijing 100101, People's Republic of China}

\author[0000-0001-9036-8543]{Wei-Yang Wang (王维扬)}
\affiliation{School of Astronomy and Space Science, University of Chinese Academy of Sciences, Beijing 100049, People's Republic of China}
\email{wywang@ucas.ac.cn}

\author[0000-0002-9725-2524]{Bing Zhang (张冰)}
\affiliation{The Nevada Center for Astrophysics, University of Nevada, Las Vegas, NV 89154, USA}
\affiliation{Department of Physics and Astronomy, University of Nevada, Las Vegas, NV 89154, USA}
\email{bing.zhang@unlv.edu}

\author[0000-0001-6475-8863]{Xuelei Chen (陈学雷)}
\affiliation{National Astronomical Observatories, Chinese Academy of Sciences, 20A Datun Road, Chaoyang District, Beijing 100101, People's Republic of China}
\affiliation{State Key Laboratory of Radio Astronomy and Technology, Beijing 100101, People's Republic of China}
\email{xuelei@cosmology.bao.ac.cn}

\author[0000-0002-9642-9682]{Jia-Wei Luo (罗佳伟)}
\affiliation{College of Physics and Hebei Key Laboratory of Photophysics Research and Application, Hebei Normal University, Shijiazhuang, Hebei 050024, People's Republic of China}
\affiliation{Shijiazhuang Key Laboratory of Astronomy and Space Science, Hebei Normal University, Shijiazhuang, Hebei 050024, People's Republic of China}

\author[0000-0002-4300-121X]{Rui Luo (罗睿)}
\affiliation{Department of Astronomy, School of Physics and Materials Science, Guangzhou University, Guangzhou 510006, People's Republic of China}

\author[0000-0001-6651-7799]{Chenhui Niu (牛晨辉)}
\affiliation{Institute of Astrophysics, Central China Normal University, Wuhan 430079, People's Republic of China }

\author[0000-0003-4721-4869]{Yuanhong Qu (屈元鸿)}
\affiliation{The Nevada Center for Astrophysics, University of Nevada, Las Vegas, NV 89154, USA}
\affiliation{Department of Physics and Astronomy, University of Nevada, Las Vegas, NV 89154, USA}

\author{Bojun Wang (王铂钧)}
\affiliation{National Astronomical Observatories, Chinese Academy of Sciences, 20A Datun Road, Chaoyang District, Beijing 100101, People's Republic of China}

\author{Fayin Wang (王发印)}
\affiliation{School of Astronomy and Space Science, Nanjing University, Nanjing 210093, People's Republic of China}
\affiliation{Key Laboratory of Modern Astronomy and Astrophysics (Nanjing University), Ministry of Education, People's Republic of China}

\author{Pei Wang (王培)}
\affiliation{National Astronomical Observatories, Chinese Academy of Sciences, 20A Datun Road, Chaoyang District, Beijing 100101, People's Republic of China}

\author{Tiancong Wang (王天聪)}
\affiliation{Department of Astronomy, Beijing Normal University, Beijing 100875, People's Republic of China}


\author{Qin Wu (吴沁)}
\affiliation{School of Astronomy and Space Science, Nanjing University, Nanjing 210093, People's Republic of China}
\affiliation{Key Laboratory of Modern Astronomy and Astrophysics (Nanjing University), Ministry of Education, People's Republic of China}

\author{Ziwei Wu (吴子为)}
\affiliation{National Astronomical Observatories, Chinese Academy of Sciences, 20A Datun Road, Chaoyang District, Beijing 100101, People's Republic of China}

\author{Jiangwei Xu (徐江伟)}
\affiliation{National Astronomical Observatories, Chinese Academy of Sciences, 20A Datun Road, Chaoyang District, Beijing 100101, People's Republic of China}

\author[0000-0001-6374-8313]{Yuan-Pei Yang (杨元培)} 
\affiliation{South-Western Institute for Astronomy Research, Yunnan University, Kunming, Yunnan 650504, People's Republic of China}
\affiliation{Purple Mountain Observatory, Chinese Academy of Sciences, Nanjing 210023, People's Republic of China}

\author[0009-0005-8586-3001]{Jun-Shuo Zhang (张钧硕)}
\affiliation{National Astronomical Observatories, Chinese Academy of Sciences, 20A Datun Road, Chaoyang District, Beijing 100101, People's Republic of China}
\affiliation{School of Astronomy and Space Science, University of Chinese Academy of Sciences, Beijing 100049, People's Republic of China} 




\begin{abstract}
Fast radio bursts (FRBs), typically highly polarized, usually have a nearly constant polarization position angle (PA) during each burst. Some bursts show significant PA variations, and one of them was claimed to have a PA variation pattern consistent with the prediction of the rotating vector model (RVM) commonly adopted to fit the PA variations in radio pulsars. We systematically study the PA evolution pattern of 1727 bursts from three active repeating FRB sources monitored by the Five-hundred-meter Aperture Spherical Telescope (FAST).
We identify 46 bursts whose PA variations are fully consistent with the RVM. However, the inferred geometrical parameters and rotation periods derived from these RVM fitting are inconsistent from each other. This suggests that the magnetosphere of the FRB central engine is constantly distorted by the FRB emitter, and the magnetic configuration is dynamically evolving.
\end{abstract}



\section{Introduction} \label{sec:intro}
FRBs are the bright millisecond radio emission \citep{2007Sci...318..777L, 2013Sci...341...53T, 2022Natur.602..585K}, which are poorly understood, especially their origins and emission mechanisms \citep{2019ARA&A..57..417C, 2019A&ARv..27....4P, 2023RvMP...95c5005Z}.
So far, hundreds of FRB sources and more than thousands of bursts have been detected \footnote{\url{https://blinkverse.zero2x.org}}.
One of these sources has been identified as a Galactic magnetar \citep{2020Natur.587...59B, 2020Natur.587...54C}, supporting the magnetar origin hypothesis.

FRBs are observed to be highly polarized. Polarization measurements serve as an effective way to investigate the radiation mechanisms and propagation effects of FRBs.
Numerous observations have demonstrated that the linear polarization position angle (PA) of FRBs exhibits a range of morphological characteristics.
For most FRBs, the PA remains constant across a burst \citep{2018Natur.553..182M, 2018ApJ...863....2G, 2019ApJ...885L..24C, 2020ApJ...896L..41C, 2021NatAs...5..594N, 2021ApJ...908L..10H, 2021MNRAS.508.5354H, 2021Natur.596..505P, 2021ApJ...911L...3P, 2022ApJ...932...98S, 2022NatAs...6..393N, 2022SciBu..67.2398F, 2023MNRAS.524.3303B, 2023ApJ...950...12M, 2023MNRAS.526.3652K}. Some others, on the other hand, show significant variations during a burst with diverse variable PA profiles \citep{2015Natur.528..523M, 2020ApJ...891L..38C, 2020Natur.586..693L, 2022RAA....22l4003J, 2022MNRAS.512.3400K, 2024arXiv241010172X, 2024arXiv241109045N, 2024ApJ...974..274F}.
The varying PA profiles favor a magnetospheric origin \citep{2017MNRAS.468.2726K,2017ApJ...836L..32Z, 2018ApJ...868...31Y, 2022MNRAS.515.2020Q,2022ApJ...927..105W,2022ApJ...925...53Z,2024ApJ...972..124Q} consistent with the line of sight (LOS) sweeping different magnetic field lines during the burst event \citep{1969ApL.....3..225R}.
The magnetospheric origin for FRB emission has been strengthened by many pieces of supporting evidence, including orthogonal jumps in PA profiles \citep{2024ApJ...972L..20N,2024NSRev..12..293J,2025arXiv250407449Q}, significant circular polarization \citep{2020MNRAS.497.3335D, 2022MNRAS.512.3400K, 2022Natur.611E..12X, 2023ApJ...955..142Z, 2024NSRev..12..293J}, characteristic S-shaped PA swings \citep{2025Natur.637...43M}, and scintillation measurements \citep{2025Natur.637...48N}.

The Rotating Vector Model (RVM; \citealt{1969ApL.....3..225R}) introduces a dipolar magnetic field that co-rotates with the neutron star to describe the observed PA swings (e.g., \citealt{2023RAA....23j4002W, 2023MNRAS.520.4582P}).
It has been widely adopted in the radio pulsar field to constrain the geometric configurations (e.g., \citealt{2001ApJ...553..341E, 2005MNRAS.364.1397J, 2008MNRAS.391.1210W, 2015MNRAS.446.3367R, 2018ApJ...856..180C}).
Recently, one CHIME-detected apparently non-repeating FRB, FRB 20221022A, revealed a PA variation that is in agreement with the RVM \citep{2025Natur.637...43M}, making a closer analogy between FRBs and pulsars and suggesting a magnetospheric origin of the FRB emission. 
Similar cases have been seen in repeating FRBs also, e.g. FRB 20180301A \citep{2020Natur.586..693L} without a systematic study. In view of the apparent interest of the CHIME burst in the community, there is a pressing need for a systematic study of the PA characteristics of an expanded sample of actively repeating FRBs, especially multiple bursts from the same source.

In this paper, we aim to investigate the morphology of PA profiles for repeating FRBs and test the validity of the RVM in FRBs. We make use of the large data set collected with the Five-hundred-meter Aperture Spherical Radio Telescope (FAST) \citep{2019SCPMA..6259502J} under the FAST FRB Key Science Project and systematically study the PA evolution profiles of a large sample of bursts.
The rest of this paper is organized as follows.
In Section \ref{sec: met}, we describe the method to fit RVM and introduce the FRB samples.
In Section \ref{sec: res}, we present the classification results of the PA morphology and the fitting results of RVM for a sub-sample of bursts.
In Section \ref{sec: dis}, the implications of our results are discussed.
Finally, conclusions are given in Section \ref{sec: con}.

\section{Method} \label{sec: met}
\subsection{PA evolution from RVM}
In the RVM picture, the PA evolution is thought to be connected to the direction of the emission region's magnetic field \citep{1969ApL.....3..225R}.
With the rotation of the dipolar field, the angle between the projected field line and the LOS also changes, resulting in PA appearing in a smooth S-shape swing.
The PA as a function of the rotation phase, $\mathrm{PA}(\phi)$, can be expressed as
\begin{equation}
\tan \left(\mathrm{PA}-\mathrm{PA}_0\right)=\frac{\sin \alpha \sin \left(\phi-\phi_0\right)}{\sin\zeta \cos (\alpha)-\cos \zeta \sin \alpha \cos \left(\phi-\phi_0\right)},
\label{eq:tanPA}
\end{equation}
where $\zeta = \alpha + \beta$ represents the angle between the spin axis and the LOS, in which $\alpha$ is the inclination angle between the rotation and magnetic axes and $\beta$ is the impact angle which is the minimum angle between the trajectory of the LOS and the magnetic axis.
The steepest gradient of $\mathrm{PA}(\phi)$ is expressed as
\begin{equation}
\left(\frac{d \mathrm{PA}}{d \phi}\right)_{\max }=\frac{\sin \alpha}{\sin \beta},
\label{eq:dPAdphi}
\end{equation}
which occurs at the phase of $\phi_{0}$.
This phase is expected to coincide with the passage of the LOS through the fiducial plane, which encompasses both the rotation and magnetic axes.
The measurement of the PA evolution over the rotation phase allows the parameter space of $\alpha$ and $\beta$ to be constrained.
The limited opening angle only allows emissions within a small range of rotation phases near the fiducial plane, therefore the constraints are not always tight.
The effective spin period $\hat P$ can be introduced to change the rotation phase $\phi$ to time $t$ so that the local configuration of the complex magnetic field can be approximated as a dipole field.
Now, the RVM would directly predict the PA evolution over $t$ and the fiducial plane corresponds to $t_{0}$:
\begin{equation}
\phi - \phi_{0} = \frac{t - t_{0}}{\hat P} \times 2 \pi.
\end{equation}
This extension may directly probe the spin period using the PA evolution.

\subsection{FRB Data Samples}
We take four burst samples from three actively repeating FRB sources monitored by FAST, including the data of FRB 20180301A, the first and second active episodes of FRB 20201124A, and the data of FRB 20220912A. The observations were made using the central beam of the L-band 19-beam receiver of FAST \citep{2020RAA....20...64J}. For convenience, these four samples are labeled by FRB 20180301A, FRB 20201124A1, FRB 20201124A2, and FRB 20220912A, respectively.
The detailed data analysis, including the pulse search, the DM optimization, and the polarimetric calibration, can be found in previous publications \citep{2020Natur.586..693L,2022Natur.611E..12X, 2022RAA....22l4003J, 2023ApJ...955..142Z}.
Since the polarization analysis requires a high signal-to-noise ratio and we focus on PA changes over time, we raise the signal-to-noise threshold for selective bursts to 50 and remove samples with fewer than 10 data points that exhibit errors less than 5 degrees.
After filtration, the number of bursts from FRB 20180301A, FRB 20201124A1, FRB 20201124A2, and FRB 20220912A was reduced to 3, 726, 536, and 462, respectively.

Here, we briefly review their main properties.
FRB 20180301A was the first repeating FRB detected with diverse PA features, including the constant PA, the smooth swing PA, and the irregular PA variations \citep{2020Natur.586..693L}.
These features strongly favor the magnetospheric origin.
During the first active episode of FRB 20201124A \citep{2022Natur.611E..12X}, the source was monitored daily for an extended period due to it high activity so that the evolution of basic properties over time was recorded.
Observations revealed irregular short-term variations in the RM of individual bursts during the first 36 days, followed by a constant RM.
More than half of the bursts exhibited circular polarization, with one burst reaching as high as 75\% degree circular polarization.
Oscillations in fractional linear and circular polarizations, as well as variations in polarization angle with wavelength, were detected in several bursts, indicating a complex and dynamically evolving magnetized environment.
The second active episode of FRB 20201124A was extremely active. The source emitted 536 bright bursts (S/N $>$ 50) in four days but was suddenly quenched afterwards.
The properties of the bursts were studied extensively in four aspects: burst morphology \citep{2022RAA....22l4001Z}, energy distribution \citep{2022RAA....22l4002Z}, polarimetry \citep{2022RAA....22l4003J}, spin-period search \citep{2022RAA....22l4004N}. The RM evolution and pulse-to-pulse RM scatter with properties are similar to the results in the first active episode.
A group of bursts from FRB 20221124A2 exhibited remarkably high circular polarization, with one instance reaching up to 90\% \citep{2024NSRev..12..293J}.
FRB 20220912A is located in a relatively clean environment \citep{2024ApJ...974..296F}, characterized by a stable Faraday rotation measure (RM) throughout its active episode.
The evolution of the circular polarization degree was also discovered in this source, suggesting the potential existence of the Faraday conversion \citep{2023ApJ...955..142Z,Wang25}.

\subsection{Likelihood Function}
Assuming a Gaussian noise model in the PA, the likelihood $\mathcal{L} \propto \exp (- \frac{\chi^2}{2})$ can be established and $\chi^2$ is expressed as
\begin{equation}
 \chi^2 (\boldsymbol{\theta}) = \sum_{i} \frac{ \left( \mathrm{PA}_{\mathrm{obs},i} - \mathrm{PA} \left( t_{i}, \boldsymbol{\theta}  \right) \right)^2 }{\sigma_{\mathrm{PA},i}^2},
\end{equation}
where $\boldsymbol{\theta}$ represents the parameters of the model.
For the constant PA model, it only has one parameter $\boldsymbol{\theta} = \{ \mathrm{PA}_{0} \}$ whereas for the RVM model $\boldsymbol{\theta} = \{\hat P,\alpha,\beta,t_{0},\mathrm{PA}_0 \}$.
$\mathrm{PA} \left( t_{i}, \boldsymbol{\theta} \right)$ is the PA profile predicted from the model with given parameters of $\boldsymbol{\theta}$ and time $t_i$.
Due to the Faraday rotation effect, linearly polarized radio emission would undergo rotation of PA proportional to the square of the wavelength when it passes through a magnetoionic medium,
\begin{equation}
    \mathrm{PA} (\lambda^2) = \mathrm{PA}_0 + \mathrm{RM} \lambda^2,
\end{equation}
where $\mathrm{PA}_0$ is the intrinsic PA at infinite frequency and RM is the rotation measure, the integrated number density of electrons multiplied by the parallel component of the magnetic field along the LOS.
Since the detailed polarization analysis of these samples have been carried out in prior studies, we can directly use the reported best-fit RM to de-rotate the de-dispersed dynamic spectrum, thereby obtaining $Q_{\mathrm{obs}, \mathrm{derot}}$ and $U_{\mathrm{obs}, \mathrm{derot}}$ directly.
$\mathrm{PA}_{\mathrm{obs},i}$ is the observed PA at time $t_{i}$, which can be determined from the de-rotated Stokes parameters $Q_{\mathrm{obs}, \mathrm{derot}}$ and $U_{\mathrm{obs}, \mathrm{derot}}$, namely
\begin{equation}
    \mathrm{PA}_{\mathrm{obs}} = \frac{1}{2} \arctan \left( \frac{U_{\mathrm{obs}, \mathrm{derot}} }{Q_{\mathrm{obs}, \mathrm{derot}}} \right),
\end{equation}
and the uncertainty of PA $\sigma_{\mathrm{PA}, i}$ can be obtained by propagating the uncertainties on the Q and U profiles,
\begin{equation}
\sigma_{\mathrm{PA}} = \frac{1}{2} \frac{Q_{\mathrm{obs}, \mathrm{derot}} U_{\mathrm{obs}, \mathrm{derot}}}{Q_{\mathrm{obs}, \mathrm{derot}}^2+U_{\mathrm{obs}, \mathrm{derot}}^2} \sqrt{ \left( \frac{\sigma_{Q_{\mathrm{obs}, \mathrm{derot}}}}{Q_{\mathrm{obs}, \mathrm{derot}}} \right)^2 + \left( \frac{\sigma_{U_{\mathrm{obs}, \mathrm{derot}}}}{U_{\mathrm{obs}, \mathrm{derot}}}\right)^2 },
\end{equation}
where $\sigma_{Q_{\mathrm{obs}, \mathrm{derot}}}$ and $\sigma_{U_{\mathrm{obs}, \mathrm{derot}}}$ represent the noise root mean square of the neighboring off-pulse region.
Since the uncertainty transfer formula may be dominated by noises in the weak part of the signal, we remove the points with errors greater than 5 degrees.

When the preparation is ready, we use the Markov chain Monte Carlo method realized by the \texttt{emcee} Python package \citep{2013PASP..125..306F} to obtain the constraints on the parameter space of $\boldsymbol{\theta}$.
To provide a fair assessment of the goodness of fit, the reduced chi-squared $\chi^2_{\nu}$ is introduced to quantify the variability of PA profiles.
It is defined as follows:
\begin{equation}
    \chi^2_{\nu} = \frac{\chi^2_{min}}{N-n},
\end{equation}
where $N - n$ represents the degree of freedom, calculated as the number of data points minus the number of the model parameters.
Following the convention established in the previous work \citep{2024ApJ...968...50P}, we choose the threshold of $\chi^2_{\nu}$ to be 5.
This threshold is significant because values falling below 5 suggest that the observational data favor the corresponding model.
Conversely, values exceeding this threshold may imply that the model is less compatible with the observations and thus is disfavored by the observational data.

\section{Results} \label{sec: res}
\subsection{Classification of the Repeating FRBs}
\begin{figure}
    \centering
    \includegraphics[width=0.45\linewidth]{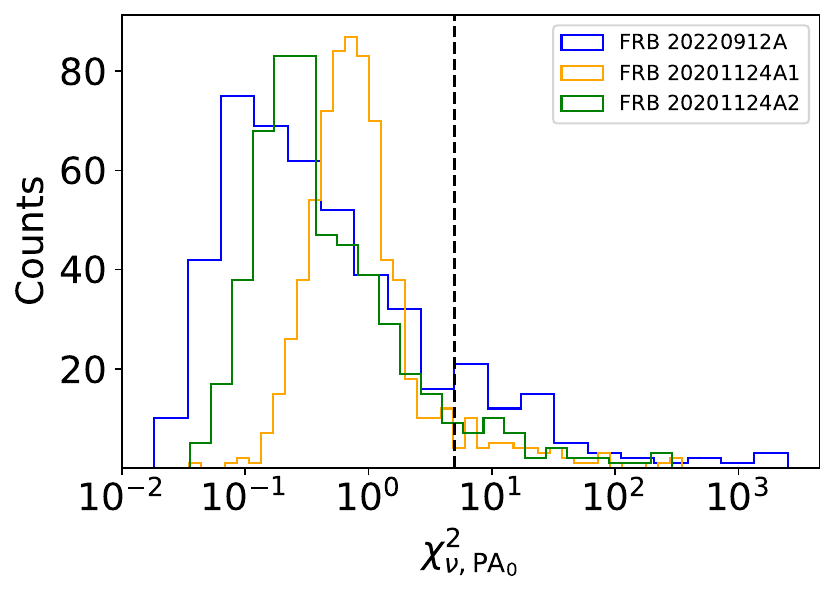}
    \includegraphics[width=0.45\linewidth]{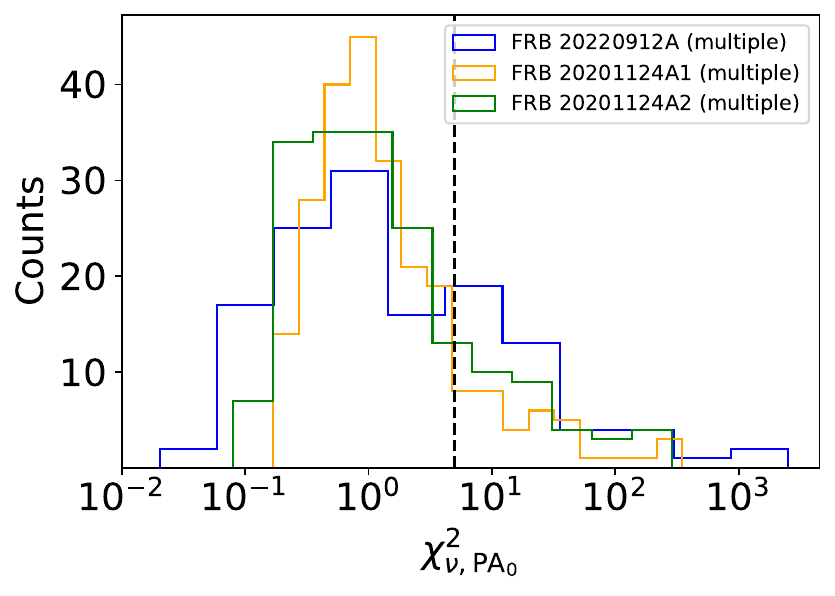}
    \caption{
    {The $\chi^2_{\nu}$ distributions of the PA profiles of FRB 20201124A1, FRB 20201124A2, and FRB 20220912A, derived from fitting a constant PA model.}
    The left panel shows the distribution for the entire sample, while the right panel focuses on the sub-sample with multiple components.
    The black vertical dotted line represents the threshold of reduced chi-squared $\chi^2_{\nu} = 5$. {$\chi^2_{\nu}$ below 5 indicates a constant PA, while $\chi^2_{\nu} = 5$ above 5 suggests a variable PA}.
    }
    \label{fig: chi}
\end{figure}

To quantify the variation of PA profiles, we performed a constant PA fit to the PA profiles of the bursts in the four samples to identify cases with PA variations. Only three bursts of FRB 20180301A meet the SNR requirement, and all of them have multiple components. We therefore ignore this small sample and focus on the three large samples of FRB 20201124A1, FRB 20201124A2, and FRB 20220912A. In Figure \ref{fig: chi}, we present the histograms of the reduced chi-squared $\chi^2_{\nu}$ of the total profiles alongside those of the multi-component profiles for each dataset.
Adhering to the predefined reduced chi-squared $\chi^2$ criteria and the number of components in the burst, we categorize all PA profiles into four distinct classes: 1. single component with a constant PA profile, 2. single component with a variable PA profile, 3. multiple components with a constant PA profile, and 4. multiple components with a variable PA profile.
The detailed classification results are summarized in Table \ref{tab1}.

The classification of the PA profiles across three data samples reveals overall consistent characteristics despite some variations in the detailed proportions.
Across all samples, the predominant type was single component profiles with a constant PA, accounting for approximately 65\%.
Specifically, 304 (65.8\%) of the PA profiles for FRB 20220912A have a single component with a constant PA, while FRB 20201124A1 and FRB 20201124A2 show slightly lower proportions, with 472 profiles (65.0\%) and 350 profiles (65.3\%), respectively.
In contrast, single component profiles with a variable PA are the rarest for all samples.
Specifically, FRB 20220912A has 24 such profiles, accounting for 5.2\%, FRB 20201124A1 has 18, representing 2.5\%, and FRB 20201124A2 has 7, making up 1.3\%.
Multiple-component profiles with a constant PA are more common in the FRB 20201124A samples, with 199 profiles (27.4\%) for the first active episode and 146 profiles (27.2\%) for the second, as compared to 94 profiles (20.3\%) for FRB 20220912A.
Multiple-component profiles with a variable PA are also rare, accounting for 37 profiles (5.1\%) in FRB 20201124A1, 33 profiles (6.2\%) in FRB 20201124A2, and 40 profiles (8.7\%) in FRB 20220912A.
Compared with the PA classification results of non-repeating FRB samples from CHIME \citep{2024ApJ...968...50P}, the overall trends are consistent.
We will further discuss these results in section \ref{dis: comparation}.

\begin{table*}
\centering
\caption{Summary of the classification of the PA profiles for the three large repeating FRB samples of FRB20201124A1, and FRB20201124A2, and FRB20220912A.
The classification results of PA profiles for CHIME's non-repeating FRBs are also compared in the table, adapted from  \cite{2024ApJ...968...50P}.}
\begin{tabular}{cccccccc}
\hline\hline
 & number of bursts & single-constant & single-variable & multiple-constant & multiple-variable \\
{FRB 20201124A1} & 726 & 472 (65.0\%) & 18 (2.5\%) & 199 (27.4\%) & 37 (5.1\%) \\
{FRB 20201124A2} & 536 & 350 (65.3\%) & 7 (1.3\%) & 146 (27.2\%) & 33 (6.2\%) \\
{FRB 20220912A} & 462 & 304 (65.8\%) & 24 (5.2\%) & 94 (20.3\%) & 40 (8.7\%) \\
\hline
Non-repeating FRBs & 88 & 50 (56.8\%) & 9 (10.2\%) & 19 (21.6\%) & 10 (11.4\%) \\
\hline\hline
\end{tabular}
\label{tab1}
\end{table*}

\subsection{RVM fitting and results}
\begin{figure}
    \centering
    \includegraphics[width=0.85\linewidth]{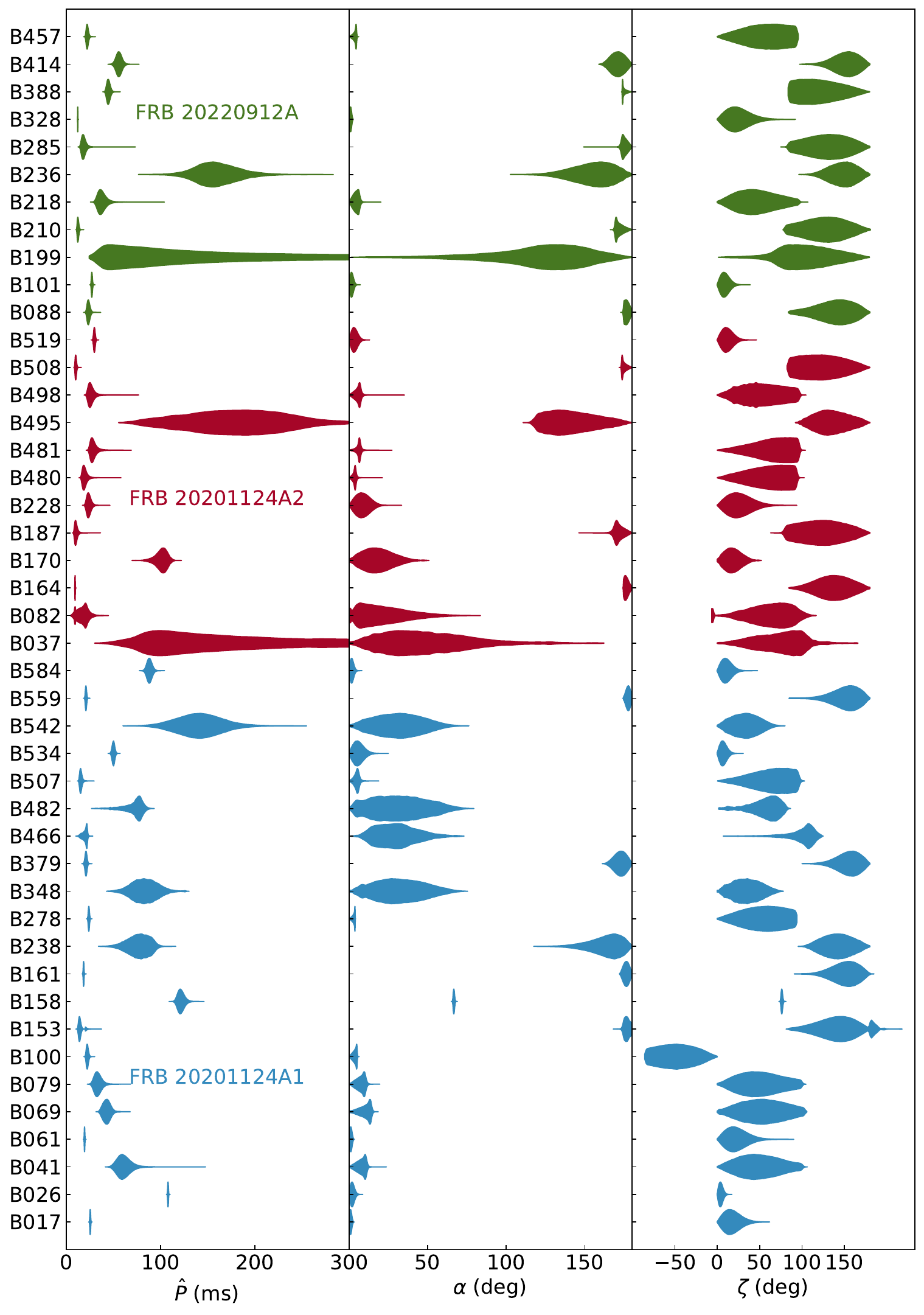}
    \caption{The violin summary of the 44 marginalized posterior distributions of the effective spin period $\hat P$, the inclination angle $\alpha$ and the angle between the rotation axis and the LOS $\zeta$ for a sub-sample of PA profiles exhibiting RVM-like variations.
    The violins from FRB 20201124A1, FRB 20201124A2, and FRB 20220912A are represented using blue, red, and green violins, respectively.
    }
    \label{fig: SUM}
\end{figure}

Among all the varying PA profiles, we selected 46 bursts that exhibit a clear RVM-like trend and fitted these profiles with the RVM to constrain the geometric parameters. The 46 PA profiles and their corresponding best-fit model curves for these bursts are illustrated in Figures \ref{fig: PAs1} and \ref{fig: PAs2}.
The marginalized posterior distributions of the effective spin period $\hat P$, the inclination angle $\alpha$, and the angle between the spin axis and LOS $\zeta$ of the 44 PA profiles from FRB 20201124A1, FRB 20201124A2, and FRB 20220912A are presented in Figure \ref{fig: SUM}.
FRB 20180301A has two bursts that can be well reproduced by RVM, but only one burst gives relatively strict constraints on geometric parameters and effective periods, so we only show their PA profiles and the best-fit model curves, and do not include them in Figure \ref{fig: SUM}.

The inferred effective periods $\hat P$ range from approximately 10 milliseconds to several hundred milliseconds.
However, previous studies \citep{2022Natur.611E..12X, 2022RAA....22l4004N, 2023ApJ...955..142Z}, which employed conventional period methods, such as Lomb-Scargle periodograms and phase folding methods, have demonstrated that the data do not support the existence of periods within this range.
We observed that many inferred effective periods $\hat P$ are relatively small, which we attribute to our selection criteria favoring bursts that exhibit significant trends over their durations.
These bursts typically possess larger duty cycles, allowing for tighter parameter constraints.
The inferred effective periods display an irregular distribution, with even bursts from the same source during the same active episode yielding significantly different inferred parameters.
The inclination angle $\alpha$ predominantly clusters near 0 or 180 degrees, suggesting that the magnetic axis is closely aligned or anti-aligned with the rotation axis.
This is consistent with the fact that most of the selected PA profiles have a rather small variation, only a few can exceed 90 degrees.
However, similar to the distribution of $\hat P$, the distribution of $\alpha$ is chaotic, with notable deviations observed in specific cases, such as B158 from FRB 20201124A1.
Additionally, we present the distribution of $\zeta$, which represents the angle between the rotation axis and the LOS.
This angle is expected to remain constant in the RVM's scenario.
However, like the case of $\hat P$ and $\alpha$, the inferred $\zeta$ also exhibits a chaotic distribution.
Within the framework of the RVM model, a chaotic distribution of $\zeta$ indicates that the emission regions of FRBs are globally stochastic within the magnetosphere.

The analysis of the PA profiles reveals that most of them exhibit a significant large duty cycle, with the smallest duty cycle identified in B495 of FRB20201124A2.
This specific PA profile has been previously examined in detail concerning its jump phenomenon \citep{2024ApJ...972L..20N}, and we have corrected the jump of the PA profile.
The profile of this event aligns well with the RVM's predictions, wherein the open field line region provides a narrow and confined site for a stable radio beam.
The opening angle constrains the duty cycle, resulting in PA profiles that typically occur within a small range near the fiducial plane.
However, contrary to the RVM's expectations, most PA profiles in our sample have duty cycles greater than 50\%.
Furthermore, these profiles do not match the behavior of RVM around the fiducial plane, as many exhibit prominent peaks.
Notably, FRB 20201124A1's B061 and B278, FRB 20201124A2's B164 and B480, along with FRB 20220912A's B101, B210, and B388, nearly display complete S-shaped curves, with duty cycles approaching unity.
FRB 20220912A's B328 demonstrates a remarkable pattern of approximately two cycles of sinusoidal oscillation, resulting in the duty cycle of roughly 2.
These facts suggest that the variation of the PA profile may have a more complicated origin than the simple RVM within the magnetic dipole framework.

\begin{figure}
    \centering
    \includegraphics[width=0.95\linewidth]{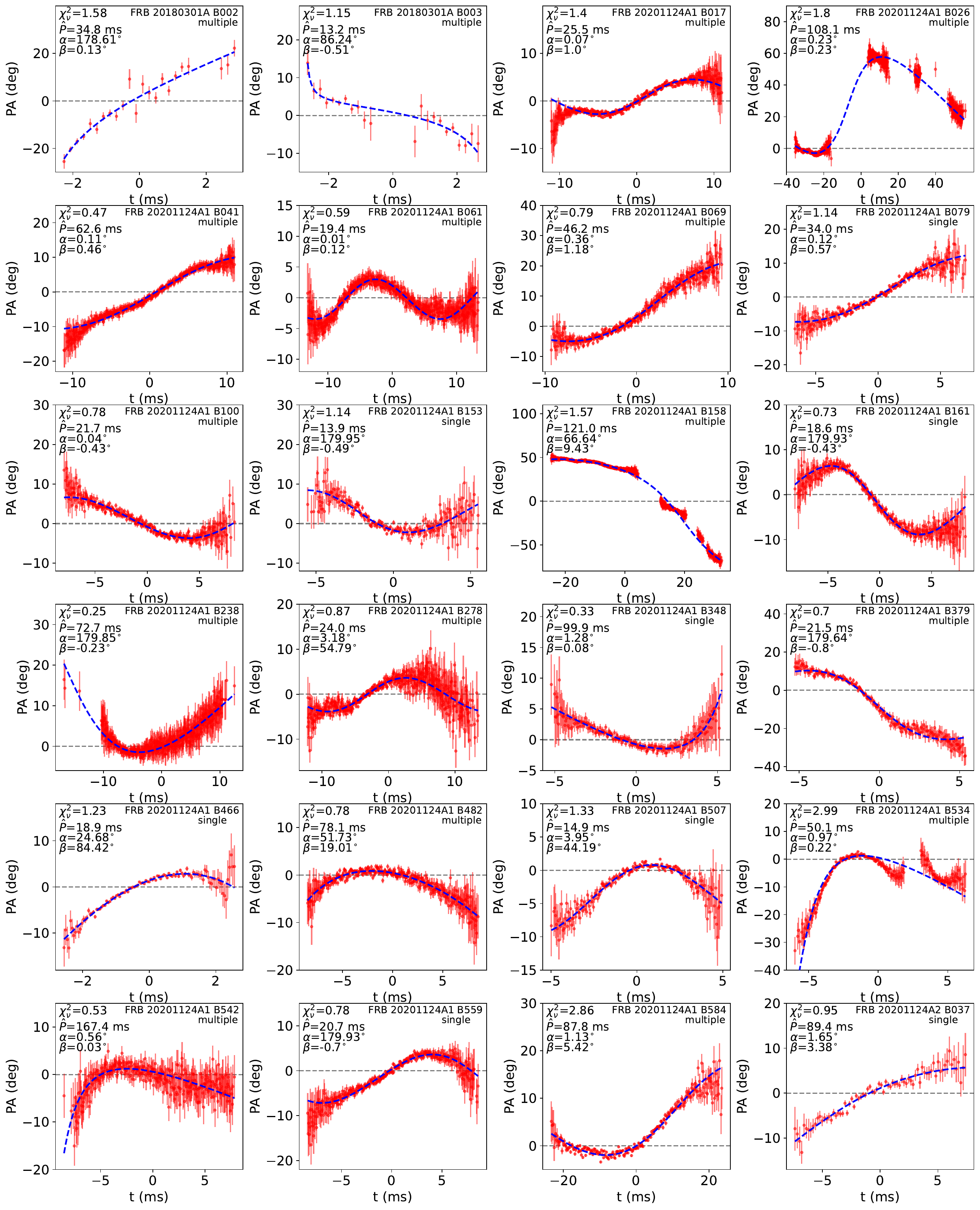}

    \caption{The measured PA profiles and the best-fit RVM curves for 46 PA profiles that exhibit RVM-like variations.
    For each panel, the corresponding sample and the identifier within that sample are labeled.
    The top left corner of each panel indicates the quality of fits $\chi^2_{\nu}$, along with the best-fit effective period $\hat P$, inclination angle $\alpha$, and the impact angle $\beta$.
    Additionally, each panel specifies whether the burst has a single or multiple components.}
    \label{fig: PAs1}
\end{figure}

\begin{figure}
    \centering
    \includegraphics[width=0.95\linewidth]{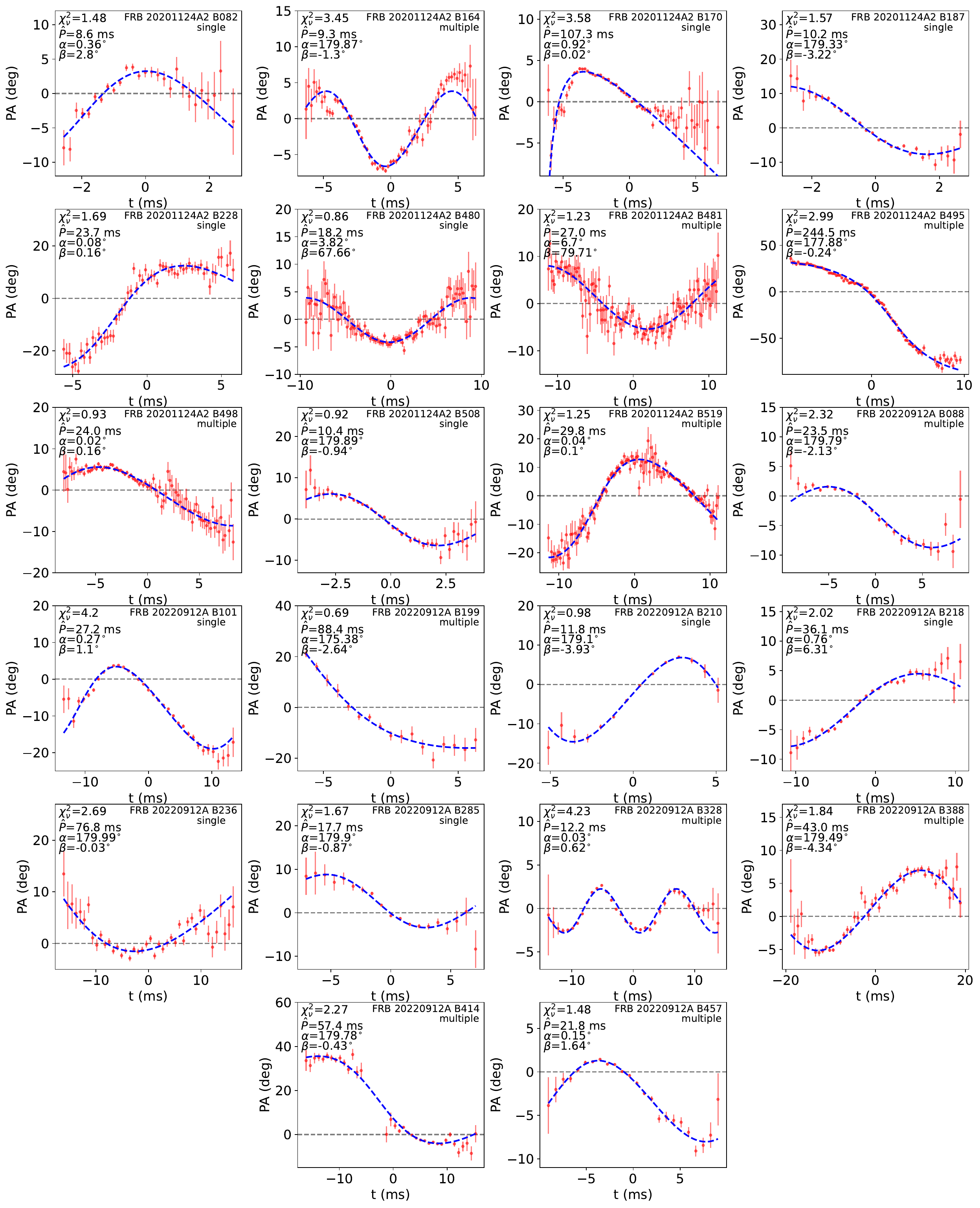}
    
    \caption{-continued.}
    \label{fig: PAs2}
\end{figure}

\section{Disscusion} \label{sec: dis}
\subsection{Comparation of the repeating and apparently non-repeating FRBs} \label{dis: comparation}
The PA classification results for FRB20220912A, FRB20201124A1, and FRB20201124A2 reveal that most repeating FRBs exhibit a flat PA profile, with only a small subset displaying significant variations.
This result is consistent with that of apparently non-repeating FRBs as observed with CHIME \citep{2024ApJ...968...50P}, but there are some differences in the specific proportions.
The classification may be affected by some factors.
The sensitivity of radio telescopes limits our ability to retrieve the PA profiles accurately.
Weak bursts often exhibit greater PA uncertainties and are potentially misclassified as the constant PA type despite their potential variations.
The definition of a burst differs somewhat between the repeating and non-repeating FRBs.
Due to the different definitions of individual bursts for repeating and non-repeating FRBs, some multiple components non-repeating FRBs may be categorized as multiple independent bursts under the criteria of repeating FRBs, potentially contributing to the inconsistency in the classification results.
Consequently, the fundamental characteristics of the PA profiles for both repeating and non-repeating FRBs can be regarded generally consistent, with the majority displaying a constant PA profile and only a minority exhibiting significant variations.
This suggests a shared underlying mechanism for repeating and apparently non-repeating FRBs.

\subsection{Origin of the PA variations}\label{dis: Origin}
The variations of PA in a good fraction of FRBs favor a magnetosphetic origin of FRBs. 
The synchrotron maser models rely on ordered magnetic fields to produce polarized emission with a constant PA \citep{2019MNRAS.485.4091M, 2020ApJ...896..142B}. While complex magnetic field configurations may realize PA variation, they disrupt the ordered magnetic fields, rendering such models less promising.
We therefore limit our discussion to these models in the following. The majority of bursts are consistent with a constant PA profile. This could arise from a few factors: a nearly aligned spin and magnetic axes (very small $\sin\alpha$ in Equation (\ref{eq:dPAdphi})), a very long spin period or a very large emission radius (very small $\sin(\phi-\phi_0)$ in Equation (\ref{eq:tanPA})). These configurations seem to be supported by the FAST observational data \citep{2025arXiv250216626L} and theoretical modeling of magnetospheric emission \citep{2017ApJ...836L..32Z,2024ApJ...972..124Q,2025ApJ...982...45B}. 
A long spin period might imply that the progenitor also had a long period, which is also similar to the long-period radio transients \citep{2025SciA...11P6351M}.

The question is how the varying PA profiles would show up in such geometric configurations. The fact that the constrained geometric parameters are inconsistent with each other from our RVM fitting suggests that the magnetosphere of the central engine is not stable, and could be dynamically evolving during the FRB burst storms. As a result, the ``effective'' magnetic axis may vary from case to case, so that under certainly configurations, given the same viewing direction, the LoS would sweep across different field lines with significant PA variations, so that the RVM cases would show up.  

The inconsistent effective periods $\hat P$ and viewing angle $\zeta$ derived from the RVM bursts also strongly suggest that the magnetospheres of the central engines (presumably magnetars) are dynamically evolving. As shown from the study of the Galactic magnetar SGR J1935+2154 \citep{2023SciA....9F6198Z}, the radio bursts seem to originate from random directions rather than focusing on a particular region such as the magnetic pole. If this is the case for cosmological FRBs, then one would expect that the observed FRBs originate from different directions. Also, unlike radio emission, FRB emission has a much higher luminosity. Together with the accompanied X-ray emission (as suggested for the Galactic FRB 20200428A, \cite{2021NatAs...5..378L,2020ApJ...898L..29M}), the ram pressure of the FRB ejecta and the radiation pressure of the radio and X-ray emissions would be much greater than the magnetic pressure, so that the magnetospheres would be significantly distorted \citep[see also][]{2020ApJ...904L..15I}. FRB emission also has the trend to straighten the field lines, making them less curved \citep{2022MNRAS.515.2020Q}. The low emission height might also make the PA curves non-RVM \citep{2025ApJ...983...43C}.
All these factors make each FRB possess a separate magnetic field configuration with different effective magnetic axes. The field lines also likely deviates from the dipolar configuration and approaching closer to a mono-polar configuration. 

\section{Conclusion} \label{sec: con}
In this paper, we have analyzed the PA profiles from FRB 20180301A, the first and second active episodes of FRB 20201124A, and FRB 20220912A, as observed by FAST, to investigate their morphological characteristics.
Utilizing previously established classification criteria, we categorized these PA profiles into four distinct types based on their variations of PA profiles and the observed features in the dynamic spectra.
Our results indicate that the majority of bursts exhibit a flat PA, with only a small fraction showing significant variations.
Notably, the classification results are consistent with the results of apparently non-repeating FRBs, suggesting that polarization characteristics may represent a common feature between these two types.

Although most of the PA profiles observed are flat, some profiles show significant variations and can be indeed reproduced by the RVM. However, the fitted parameters, including effective period $\hat P$, inclination angle $\alpha$, and $\zeta$, display an inconsistent behavior.
These results do not support the notion that PA variations are driven by a stable magnetic rotator. Instead, they indicate that FRBs likely originate from the dynamically evolving magnetosphere. This is expected, because the magnetosphere of the FRB central engine is likely to be constantly distorted by the FRB emitter, whose ram pressure (and the FRB, X-ray emission pressure) greatly exceeds the magnetic pressure in the magnetosphere.

\begin{acknowledgments}
We are grateful to the referee. This work made use of data from the FAST FRB Key Science Project.
X.-H. L. acknowledges support from the National SKA Program of China （Nos. 2022SKA0110100 and 2022SKA0110101) and NSFC (grant No. 12361141814).
W.-Y. W. acknowledges support from the NSFC (No.12261141690 and No.12403058), the National SKA Program of China (No. 2020SKA0120100), and the Strategic Priority Research Program of the CAS (No. XDB0550300).
X.-L. C. acknowledges support from the NSFC (grant No. 12361141814).
J.-W. L. acknowledges support from the NSFC (No. 12403100).
\end{acknowledgments}

%

\vspace{5mm}


\software{numpy \citep{harris2020array}, scipy \citep{2020SciPy-NMeth}, matplotlib \citep{Hunter:2007}, emcee \citep{2013PASP..125..306F}
          }




\bibliography{sample631}{}
\bibliographystyle{aasjournal}


\end{CJK*}
\end{document}